\documentclass[10 pt, conference]{ieeeconf}
\IEEEoverridecommandlockouts                              
\usepackage{graphicx}
\usepackage{amsmath,amssymb,amsfonts}
\usepackage{subcaption}
\usepackage{color}
\usepackage{placeins}
\newtheorem{theorem}{\textbf{Definition}}

\usepackage{breakurl}
\usepackage{fixltx2e}

\begin{document}
\color{red}
\onecolumn
\huge 
\begin{center}
2019 IEEE International Conference on Blockchain (Blockchain)
\end{center}
\vskip 1.5in
\Large 
978-1-7281-4693-5/19/\$31.00 ©2019 IEEE

DOI 10.1109/Blockchain.2019.00011
\vskip 1.5in
\Large 
"\textcopyright \textcopyright 2019 IEEE. Personal use of this material is permitted. Permission from IEEE must be obtained for all other uses, in any current or future media, including reprinting/republishing this material for advertising or promotional purposes, creating new collective works, for resale or redistribution to servers or lists, or reuse of any copyrighted component of this work in other works."

\color{black}
\twocolumn

\title{\LARGE \bf
Cascading Machine Learning to Attack Bitcoin Anonymity
}

\author{
\authorblockN{Francesco Zola\authorrefmark{1}, Maria Eguimendia\authorrefmark{1}, Jan Lukas Bruse\authorrefmark{1}, Raul Orduna Urrutia\authorrefmark{1}}
\authorblockA{\authorrefmark{1} Dept. Data Intelligence for Energy and Industrial Processes, Vicomtech\\ Paseo Mikeletegi 57, 20009 Donostia/San Sebastian, Spain\\ \{fzola, meguimendia, jbruse, rorduna\}@vicomtech.org}
}

\maketitle
\thispagestyle{empty}
\pagestyle{empty}

%%%%%%%%%%%%%%%%%%%%%%%%%%%%%%%%%%%%%%%%%%%%%%%%%%%%%%%%%%%%%%%%%%%%%%%%%%%%%%%%
\begin{abstract}
Bitcoin is a decentralized, pseudonymous cryptocurrency that is one of the most used digital assets to date. Its unregulated nature and inherent anonymity of users have led to a dramatic increase in its use for illicit activities. This calls for the development of novel methods capable of characterizing different entities in the Bitcoin network.

In this paper, a method to attack Bitcoin anonymity is presented, leveraging a novel cascading machine learning approach that requires only a few features directly extracted from Bitcoin blockchain data. Cascading, used to enrich entities information with data from previous classifications, led to considerably improved multi-class classification performance with excellent values of Precision close to 1.0 for each considered class. Final models were implemented and compared using different machine learning models and showed significantly higher accuracy compared to their baseline implementation. Our approach can contribute to the development of effective tools for Bitcoin entity characterization, which may assist in uncovering illegal activities.

\end{abstract}

\IEEEoverridecommandlockouts

%%%%%%%%%%%%%%%%%%%%%%%%%%%%%%%%%%%%%%%%%%%%%%%%%%%%%%%%%%%%%%%%%%%%%%%%%%%%%%%%
\section{Introduction}
Bitcoin was born in 2009 and since then its value and popularity has been rapidly increasing until its current state, in which it is the most used, assessed and priced cryptocurrency of all. Bitcoin is a pure peer-to-peer cryptocurrency \cite{nakamoto2008bitcoin} where all transactions are stored in a public shared ledger called blockchain that cannot be manipulated or changed \cite{crosby2016blockchain}. Bitcoin is decentralized, which means that it is not controlled by any financial institution but it is regulated by everyone in the Bitcoin network: its blockchain architecture maintains the system without ambiguity \cite{narayanan2016bitcoin}. 

While transactions within the Bitcoin network are openly available, Bitcoin user identity is non-transparent and protected by anonymity. This circumstance, combined with the unregulated nature of the Bitcoin market, has brought a lot of new actors to the Bitcoin network using cryptocurrency for illicit operations. Approximately one-quarter of Bitcoin users and half of all Bitcoin transactions are associated with illegal activity \cite{foley2018sex}, accounting for an annual amount of around \$72 billion (report 2018).

Conventional law-enforcement strategies tackling illegal financial operations such as money laundering or transactions funding criminal operations are typically based on complete knowledge of each actor's identity, while details about financial transactions are controlled by banks and thus unknown \cite{moser2014towards}. Within the Bitcoin network, these circumstances are reversed - incomplete knowledge of identities restricts traceability and transparency of operations, in turn promoting further increase of illegal activities. This calls for novel methods to attack anonymity within the Bitcoin network, aiming to uncover Bitcoin entity categories.

Among the most active categories of entities is the exchange, which represents a digital marketplace where traders can buy and sell cryptocurrencies using different fiat (money made legal tender by a government decree) or other digital currencies. Exchanges thus constitute the "front and exit doors" to the cryptocurrency world and are ideal to hide illicit operations, as documented in \cite{moore2013beware}. Another category is the darknet market. These markets are e-commerce platforms where users can find drugs, weapons and any kind of goods or services that are illegal in most countries. These cryptomarkets use electronic currencies to facilitate licit and illicit transactions among their users \cite{christin2013traveling}. Further, so-called mixers represent services that allow users to obscure operations, as presented in \cite{moser2013anonymity}. At the same time mixed transactions increase the privacy of the users, and they can be used for money laundering of illegal funds. 

Being able to classify anonymous Bitcoin entities according to such categories would increase transparency and would facilitate linking blockchain information with real actors to uncover illegal activities. Current techniques attacking anonymity often try to cluster addresses and apply heuristic assumptions combined with labelled data from external sources like markets, forums or social media in order to determine address owners in the real world \cite{meiklejohn2013fistful}. However, gathering external data and combining them with Bitcoin information is tedious and could be limited due to privacy restrictions. This motivates the implementation of a model able to characterize different behaviours in the Bitcoin network by analyzing the pure blockchain information only; by extracting transactions and by recognizing patterns using machine learning approaches.

In this paper, we present a novel approach to decrease Bitcoin anonymity based on a cascading machine learning model, using entity, address and motifs data as inputs. We apply a "cascade" of classifiers, performing a first entity classification based on address, 1\_motif, and 2\_motif data, which is then used as input for a second classification step, which combines those classification results with entity information from the blockchain. Notably, our approach only requires a few features that can be directly extracted from Bitcoin blockchain data.

In order to compare benefits and limits of the proposed approach, two experiments are presented: firstly, a simple classifier is trained based on pure entity information gathered from the blockchain. In the second experiment, a final classifier is trained using the enriched data set generated by our cascading approach. We aimed to detect six different types of Bitcoin entity behaviours. Overall, three classifier models are tested and compared: Adaboost, Random Forest and Gradient Boosting.

The rest of the paper is organized as follows. Section~\ref{sec:related} describes the related work. After that, Section~\ref{sec:graph} presents the graph model used and Section~\ref{sec:data} shows an overview of the used data sets. Section~\ref{sec:machine} describes the implemented machine learning models and Section~\ref{sec:result} presents the obtained results. Finally, in Section~\ref{sec:conclusions}, we draw conclusions and provide guidelines for future work.

\section{Related work}\label{sec:related}

User anonymity has probably been the key factor for the success of cryptocurrencies and has promoted illegal activities within the Bitcoin network. Yet, several studies determine that current measures adopted by the Bitcoin protocol are not sufficient to protect the privacy of its users \cite{meiklejohn2015privacy}, \cite{androulaki2013evaluating}, opening up possibilities to attack Bitcoin anonymity. One of the first transaction analysis is documented in \cite{ron2013quantitative} where typical behavior of Bitcoin users are detected based on how they spend cryptocurrencies, how they keep the balance in their accounts, and how they move Bitcoins between their various accounts. Herrera-Joancomart{\'\i} \cite{herrera2015research} presents a review on Bitcoin anonymity, concluding that anonymity can be reduced by address clustering or by gathering information from various peer-to-peer networks. This technique is also advocated in \cite{koshy2014analysis}, where conservative constraints (patterns) are applied for address clustering, and in \cite{liao2016behind} where information gathered from online forums is used to characterize the CryptoLocker, a family of ransomware. Similarly, in \cite{fleder2015bitcoin}, information scraped from online forums and social media is determinant to simulate an attacker and to summarize activity of both known and unknown Bitcoin users. In \cite{biryukov2014deanonymisation}, a generic method to deanonymize a significant fraction of Bitcoin users by correlating their pseudonyms with public IP addresses is described. Reid et al. \cite{reid2013analysis} demonstrates how it is possible to associate many public-keys with each other, using a map of the topological network and external identifying information in order to investigate a large theft of Bitcoins.

Several recent studies have exploited machine learning algorithms for Bitcoin analysis. In \cite{hirshman2013unsupervised}, an unsupervised learning model is presented with the aim to identify atypical transactions related to money laundering. Monamo et al. \cite{monamo2016unsupervised} introduce a k-means classifier for object clustering and fraudulent activity detection in Bitcoin transactions. Another study on detection of anomalous behavior, suspicious users and transactions is presented in \cite{pham2016anomaly}, where three unsupervised learning methods are applied to two graphs generated by the Bitcoin transaction network. Further, a supervised machine learning algorithm is used by \cite{harlev2018breaking} to uncover Bitcoin anonymity using a method for predicting the type of yet-unidentified entities. In \cite{bartoletti2018data}, data mining techniques are used to implement and train a classifier to identify Ponzi schemes in the Bitcoin blockchain and in \cite{mcnally2018predicting} a Bayesian optimized recurrent neural network (RNN) and a Long Short Term Memory (LSTM) are implemented to predict the direction of Bitcoin price in USD.

Recently, an interesting approach is given in \cite{ranshous2017exchange}, where the concept of motifs is introduced to blockchain analysis. Authors performed an analysis of the transaction directed hypergraph in order to identify several distinct statistical properties of exchange addresses. They were able to predict if an address is owned by an exchange with $>80\%$ accuracy. The introduction of hypergraphs (or dirhypergraphs) proved beneficial due to their significant advantages over a complex graph structure typically derived from Bitcoin networks. In \cite{jourdan2018characterizing}, the motif concept is further developed and is combined with multiple features (entity, address, temporal, centrality) to obtain a comprehensive entity classification into five categories: Exchange, Service, Gambling, Mining Pool and DarkNet marketplace. Using a total of $315$ features, a global accuracy of $0.92$ could be achieved.

Inspired by the good classification results presented in \cite{jourdan2018characterizing}, we present here a novel machine-learning-based approach to attack Bitcoin anonymity, making use of motifs as introduced by Ranshous et al. and allowing for multi-class classification of Bitcoin entities as in \cite{jourdan2018characterizing}, yet aiming to provide a straightforward methodology that relies on fewer, well-defined features. To achieve this, we introduce a novel cascading machine learning model for Bitcoin data analysis. The main idea is to implement a cascade of classifiers, so that outgoing classification results can be joined and can be used to enrich a final classification.

\section{Graph Model}\label{sec:graph}
\subsection{Blockchain Graph Model}\label{blockchain_model}
\begin{figure}[!htbp]
  \centering
    \includegraphics[scale=0.5]{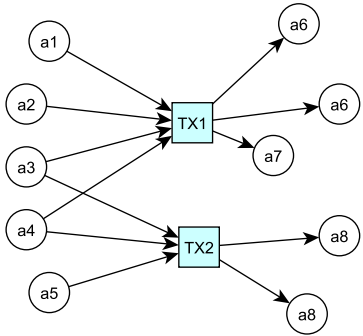}
    \caption{\textit{Example of address-transaction graph}}
    \label{fig:address}
\end{figure}

Bitcoin transactions have a natural graph structure, with a fundamental example being the address-transaction graph (Figure \ref{fig:address}). This graph is directly obtained by using the information gathered from the blockchain and provides an estimation of the flow of Bitcoins linking public key addresses over time. The vertices represent the addresses $(a\textsubscript{1},a\textsubscript{2},...,a\textsubscript{N})$ and the transactions $(tx\textsubscript{1}, tx\textsubscript{2},..., tx\textsubscript{M})$. The directed edges (arrows) between entities and transactions indicate the incoming relations, while directed edges between transactions and entities correspond to outgoing relations. Each directed edge can also include additional features such as values, time-stamps, etc.
 
\begin{figure}[!htbp]
  \centering
    \includegraphics[scale=0.5]{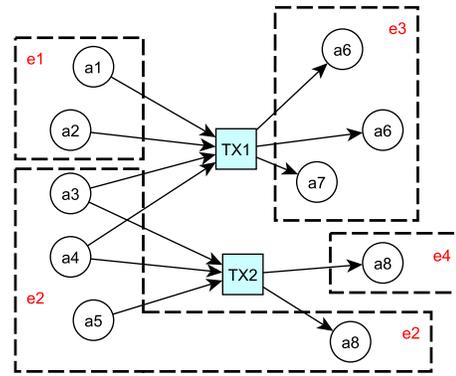}
    \caption{\textit{Example of entity-transaction graph obtained by address clustering}}
    \label{fig:entity}
\end{figure}

To improve anonymity in the network, users are encouraged to generate a new Bitcoin address for each new transaction, which is a common advice for the correct usage of Bitcoin\footnote{https://bitcoin.org/en/protect-your-privacy}. Due to this procedure, several addresses belong to the same logical user, so that a simplification is possible by introducing the concept of \textit{entities}. An entity is defined as person or organization that controls or can control multiple public key addresses. This definition allows us to transform the address-transaction graph into the entity-transaction graph (Figure \ref{fig:entity}).

The new graph is obtained by grouping addresses belonging to the same user into entities (address clustering). This operation is not intuitive, however several heuristic properties have already been presented with the aim to help the clusterization process, for example in \cite{androulaki2013evaluating}, \cite{koshy2014analysis} and \cite{ermilov2017automatic}. In the obtained graph, vertices represent the entities $(e\textsubscript{1},e\textsubscript{2},...,e\textsubscript{K})$ and the transactions $(tx\textsubscript{1},tx\textsubscript{2},...,tx\textsubscript{M})$. Similar to the address-transaction graph, directed edges between entities and transactions indicate the incoming relations, while directed edges between transactions and entities correspond to outgoing relations. The entity-transaction graph (\ref{fig:entity}) summarizes the network well and constitutes an easily understandable representation of the money flow within the network.

\subsection{Motifs Graph Model}\label{motifs_model}
Graph motifs were introduced in \cite{lacroix2006motif} and were motivated by applications in bioinformatics, specifically in metabolic network analysis. However, as shown in Section \ref{sec:related}, prior studies such as \cite{ranshous2017exchange} have introduced the concept of motifs to Bitcoin analysis. In this paper, a definition of $N\_motif$ is used, starting from the generalized concept introduced in \cite{jourdan2018characterizing}.

\begin{theorem}
A $N\_motif$ is a path from the entity-transaction graph with length $2N$ that starts and ends with an entity.
Let $(e\textsubscript{1},..,e\textsubscript{M}) \in{E}$ be a class of entities and $(t\textsubscript{1},..,t\textsubscript{N}) \in{T}$ be a class of transactions, with $M\leq{N+1}$, then:
\[N\_motif ={(e\textsubscript{1},t\textsubscript{1},...,t\textsubscript{N},e\textsubscript{M})}\]
in which at least one output from each transaction must be an input to the next transaction.

The term \textit{branch} is used here to refer to a path in the motif graph that begins and ends with an entity passing through exactly one transaction. If a single branch of the graph has the same entity as input and output ($e\textsubscript{j}=e\textsubscript{j+1}$), the branch is called Direct Loop, otherwise it is called Direct Distinct.

From the motif definition it is clear that all transactions are ordered in time, which means that $\tau(t\textsubscript{1})<\tau(t\textsubscript{2})<..<\tau(t\textsubscript{N})$, where $\tau$ represents a transaction time.
\end{theorem}

Here, we use the $1\_motif$ and $2\_motif$ concepts. The $1\_motif$ represents the relation between two entities (at least one distinct), while the $2\_motif$ is the relation between three entities (at least one distinct) involved in two consecutive transactions.

\section{Data Overview}\label{sec:data}
We considered the whole Bitcoin blockchain data created until February $5$th $2019$, $08$:$13$:$31$ AM, corresponding to $561$,$620$ blocks, which contain about $380$,$000$,$000$ transactions and involve more than $1$,$000$,$000$,$000$ addresses. This data was then combined with information available on the WalletExplorer\footnote{https://www.walletexplorer.com/}, a benchmark platform for entities detection, which represents a collection of information about different known entities that have been detected until today. The data set is thus composed of 311 different samples, divided into six classes (see Table \ref{tab:walexpl}):

\begin{itemize}
\item \textit{Exchange}: entities that allow their customers to trade
fiat currencies for Bitcoins (or vice versa)

\item \textit{Service}: entities that offer Bitcoin payment methods as solutions to their business (financial services, trading, lending, etc.)

\item \textit{Gambling}: entities that offer gambling services (casino, betting, roulette, etc.)

\item \textit{Mining Pool}: entities composed of a group of miners that work together sharing their resources in order to reduce the volatility of their returns

\item \textit{Mixer}: entities that offer a service to obscure the traceability of their clients' transactions

\item \textit{Marketplace}: entities allowing to buy any kind of goods or services that are illegal in most countries paying with Bitcoin
\end{itemize}

\begin{table}[!htbp]
\centering
\begin{tabular}{lcccc}
\hline
Class & \ Abbreviation & \# Entities & \# Address & \% Address \\ \hline
            \textit{Exchange}      &Ex            & 137               & 9,943,512                   & 61.63                   \\
            \textit{Gambling}       &Gmb           & 76               & 3,054,238                   & 18.93                  \\
            \textit{Marketplace}    &Mrk                & 20               & 2,349,210                  & 14.56                   \\
            \textit{Mining Pool}    &Pool               & 25               & 76,104                   & 0.47                 \\  \textit{Mixer}          &Mxr            & 37               & 475,714                   & 2.95                   \\   \textit{Service}        &Serv              & 16               & 235,629                   & 1.46                   \\\hline
\textbf{Total}                & & \textbf{311}               & \textbf{16,134,407}                  & \textbf{100}               \\\hline
\end{tabular}
    \caption{\textit{Overview of WalletExplorer data used for this study}}
    \label{tab:walexpl}
\end{table}

As shown in Table \ref{tab:walexpl}, the \textit{Exchange} is the top class represented by more than $60\%$ of samples, while the \textit{Mining Pool} class is the least represented with just $0.47\%$ (even though it has more distinct entities than the \textit{Marketplace} and the \textit{Service}).

\begin{figure}[!htbp]
  \centering
    \includegraphics[width=\linewidth]{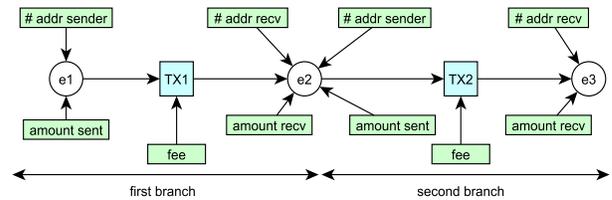}
    \caption{\textit{2\_motif representation with extracted features highlighted}}
    \label{fig:2motifs}
\end{figure}

Cross-references between Bitcoin blockchain data and labelled data from the WalletExplorer allow us to re-size the original data set by removing all the unlabelled and unusable data. As such, we focus our analysis on known entities only. From this new data set, four dataframes (2-dimensional labelled data structure or data table with samples as rows and extracted features as columns) were extracted for the proposed analysis:

\begin{itemize}
\item \textit{Entity dataframe} contains all features related to an entity that can be directly extracted from the blockchain. They are: the amount of BTC received/sent, the balance of the entity, the number of transactions in which this entity is the receiver/sender, and the number of addresses belonging to this entity used for receiving/sending money. (This dataframe was composed of $311$ samples and $7$ features)

\item \textit{Address dataframe} contains all features related to Bitcoin addresses. Features are: the number of transactions in which a certain address is detected such as receiver/sender, the amount of BTC received/sent from/to this address, the balance, uniqueness (if this address is just used in one transaction) and siblings. (This dataframe was composed of $16$,$134$,$407$ samples and $7$ features)

\item \textit{1\_motif dataframe} contains the information directly extracted from the $1\_motif$ graph. In this case, each row contains: the amount received/sent in the transaction, number of distinct addresses used for receiving/sending money, number of similar received/sent transactions between the entities in the branch, the fee, and if the branch realizes a Direct Loop or Direct Distinct path. (This dataframe was composed of $58$,$076$,$963$ samples and $9$ features)

\item \textit{2\_motif dataframe} contains information gathered from the $2\_motif$ graph. The features analyzed are: the number of addresses as input/output for the first and second path in $2\_motif$ graph, the amount received/sent in the first and second branch, the fee of both considered transactions, number of similar sent transactions between the entities in the first and second branch, Direct Loop or Direct Distinct path for the first and the second branch and Direct Loop or Direct Distinct path considering the whole $2\_motif$ path, see Figure~\ref{fig:2motifs}. (This dataframe was composed of $83$,$443$,$055$ samples and $18$ features)
\end{itemize}

\section{Machine learning}\label{sec:machine}
\subsection{Classifier Models}\label{classifier}
\begin{figure}[!htbp]
  \centering
    \includegraphics[width=\linewidth]{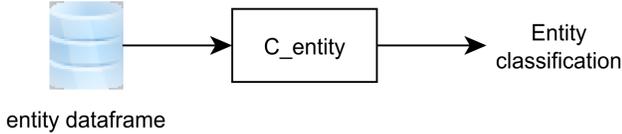}
    \caption{\textit{First experiment: simple entities classifier}}
    \label{fig:classifier_ent}
\end{figure}

To demonstrate benefits and limits of our approach, we conducted two different experiments. Firstly, we created a simple classifier, called \textit{C\_entity} (Figure \ref{fig:classifier_ent}), merely based on the samples stored in the entity dataframe, containing (seven) entity-related features that can be directly extracted from the blockchain. This classifier was evaluated via a cross-validation process (see Section \ref{evaluation}). Results from cross-validation were considered as our baseline classification. The simple classifier was implemented in three versions applying Adaboost, Random Forest and Gradient Boosting models as those previously yielded good classification results for Bitcoin data \cite{ranshous2017exchange}.

\begin{figure}[!htbp]
  \centering
    \includegraphics[width=\linewidth]{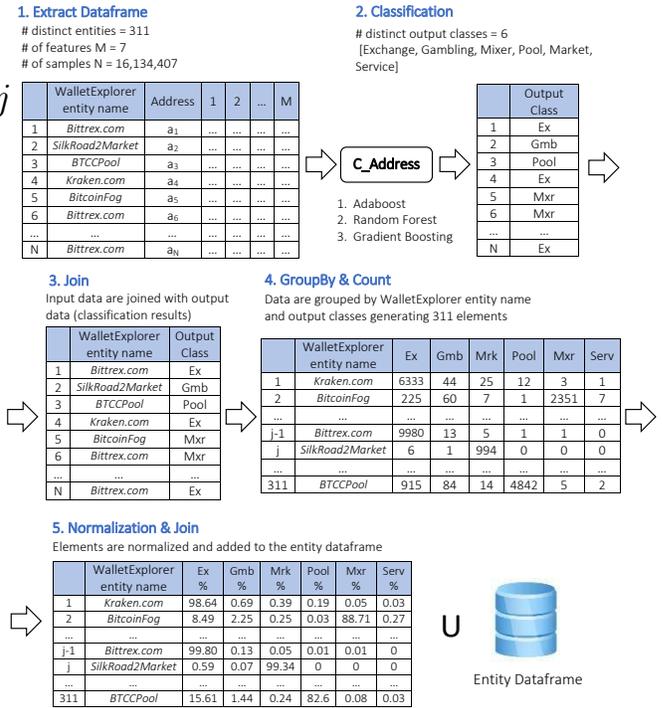}
    \caption{\textit{Steps to create the enriched entity dataframe applied to an example address dataframe}}
    \label{fig:enrichment}
\end{figure}

In the second experiment, prior to entity classification according to the six classes (Table \ref{tab:walexpl}), we built three separate classifiers, based on the additionally available address, 1\_motif, and 2\_motif dataframes and their respective features ($7 + 9 + 18 = 34$ features). Outgoing information from these classifications was processed, as shown in Figure \ref{fig:enrichment}, in order to create a set of six new features for each classifier, which were then used to enrich (extend) the entity dataframe. Finally, a new classifier \textit{C\_final} was generated to obtain final entity classification based on this enriched entity dataframe and its $25$ features ($7$ belonged to the entity dataframe and $6$x$3$ were generated from the three classifiers \textit{C\_address}, \textit{C\_motif1}, \textit{C\_motif2}). With this cascading approach, new entity-related characteristics were added to the entity dataframe, ultimately improving the classification as demonstrated in the following sections.

\begin{figure}[!htbp]
  \centering
    \includegraphics[width=\linewidth]{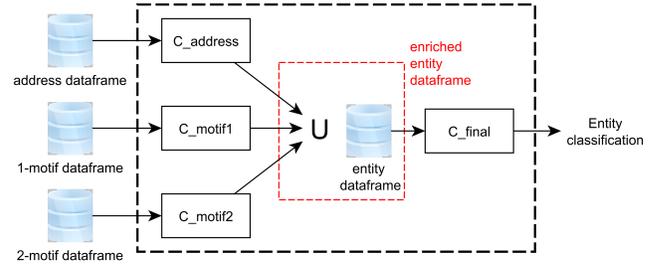}
    \caption{\textit{Second experiment: cascading entities classifiers}}
    \label{fig:classifier}
\end{figure}

The first step was to split the address, 1\_motif and 2\_motif dataframes into two parts called A-data set (for training) and B-data set (for testing) with a proportion of $70$/$30$. The A-data set was used to compute cross-validation of the three \textit{C\_address}, \textit{C\_motif1}, \textit{C\_motif2} classifier models (Figure \ref{fig:classifier}). After that, the B-data set was used as input for the trained classifiers \textit{C\_address}, \textit{C\_motif1}, \textit{C\_motif2} in order to obtain classification results based on completely new, unseen data.

Classification results essentially assign one of the six possible output classes to each entry in the input dataframe. As each entry has its original (ground truth) label obtained from the WalletExplorer, we can join input label and computed output class and perform a group-by and count operation as illustrated in Figure \ref{fig:enrichment}: we count how many times a sample belonging to a particular entity has been detected in each of the considered classes. This value is then normalized as indicated in the following formula:

$$\forall \xi \in E \qquad \frac{\parallel P\textsubscript{$\xi\vert$ j}\parallel}{\sum_{i=1}^{N}\parallel P\textsubscript{$\xi\vert$ i}\parallel }*100 \qquad with \quad j\in N$$

where $E$ is the entities set and $N$ represents the number of considered classes ($N=6$ in this study). The term $\parallel P\textsubscript{$\xi\vert$ j}\parallel$ represents how many times a sample originally labelled with entity $\xi$ generates a prediction belonging to the class j, while the term $\sum_{i=1}^{N}\parallel P\textsubscript{$\xi\vert$ i}\parallel$ counts all the predictions generated from samples with labelled input belonging to entity $\xi$.

These normalized values form a dataframe containing $311$ samples (one for each known entity as in the entity dataframe) and six new features, representing the percentage of being classified as belonging to one of the six classes. These features were added to the entity dataframe for data enrichment, constituting our cascading machine learning system. The elements of the enriched entity dataframe were used to implement and evaluate the final classifier, called \textit{C\_final}, and a cross-validation process (Section \ref{evaluation}) was applied to compute its performance.

To allow for better comparison between experiments, we implemented all classifier models \textit{C\_address}, \textit{C\_motif1}, \textit{C\_motif2} and \textit{C\_final} with Adaboost, Random Forest and Gradient Boosting models. Specifically, all Adaboost classifiers were generated with the number of estimators set to $50$ and the learning rate set to $1$. All Random Forest models were implemented with the number of estimators set to $10$, a Gini function to measure the quality of the split and without a maximum depth of the tree. All Gradient Boosting models were implemented with the number of estimators set to $100$, the learning rate set to $0.1$ and the maximum depth for limiting the number of nodes set to $3$.

\subsection{Evaluation Metrics}\label{evaluation}
All classification models were evaluated by extracting and comparing classification metrics via a cross-validation process. The goal of cross-validation is to analyze the prediction capabilities of the model in order to detect problems such as over-fitting or selection bias \cite{cawley2010over}. Here, we used stratified K-fold cross-validation, with a value of K equal to $5$. This method involves dividing the whole data set into K equal partitions or folds.

Each fold is composed of data ensuring a good representative sample of the whole population by keeping the same proportion of classes present in the original data set (stratification). Then, K-1 folds are used to train the model and the one left-out fold is used to evaluate the predictions obtained by the trained model. The entire process is repeated K times, until each fold has been left out once, testing all possible combinations. During this process, the following metrics were computed: 

\begin{itemize}
\item \textit{Accuracy} or \textit{Score} is defined as the number of correct predictions divided by the total number of predictions and is given as percentage
\item \textit{Precision} is the number of positive predictions divided by the total number of the positive class values predicted. It represents a measure of a classifier's exactness given as a value between $0$ and $1$, with $1$ relating to high precision
\item \textit{Recall} represents a measure of a classifier's completeness given as a value between $0$ and $1$
\item \textit{F\textsubscript{1}-score} is the harmonic mean of Precision and Recall. It takes values between $0$ and $1$, with $1$ relating to perfect Precision and Recall
\[F\textsubscript{1} score= 2* \frac{Precision * Recall}{Precision+Recall}\]
\item \textit{Matthews Correlation Coefficient (MCC)} is a metric yielding easy comparison
with respect to a random baseline, suitable for unbalanced classes. It takes values between $-1$ and $+1$. A coefficient of $+1$ represents a perfect prediction, $0$ an average random prediction and $-1$ an inverse prediction. As shown in \cite{gorodkin2004comparing}, let $K$ be the number of classes and  $C$ be a confusion matrix with dims $K\times K$, the $MCC$ can be calculated as:

\[MCC\_part1 ={\sqrt{\sum_{k}(\sum_{l}C\textsubscript{kl})(\sum_{f,g\vert f\neq{g}}C\textsubscript{gf})}}\]

\[MCC\_part2 ={\sqrt{\sum_{k}(\sum_{l}C\textsubscript{lk}) (\sum_{f,g\vert f\neq{g}}C\textsubscript{fg})}}\]

\[MCC = \frac{\sum_{k}\sum_{l}\sum_{m}C\textsubscript{kk}C\textsubscript{lm}-C\textsubscript{kl}C\textsubscript{mk}}{MCC\_part1*MCC\_part2}\]
\end{itemize}

In Section \ref{sec:result} results for the baseline model (\textit{C\_entity}) and for the final model obtained after cross-validation using the enriched dataframe (\textit{C\_final}) are presented and compared. We report global metric values for Accuracy/Score and \textit{MCC} averaged over the K=$5$ cross-validation runs and per-class values for Precision, Recall and F1-score when evaluating the final models.

\subsection{Hardware and Software Configuration}\label{configuration}
All analyses were run on a cluster of three virtual machines, each one with $16$ CPUs Intel(R) Xeon(R) Silver $4114$ CPU @ $2.20$ GHz, $64$ GB RAM DDR4 memory with $2,666$ MHz, and $500$ GB of Hard Disk SATA. Apache Spark\footnote{https://spark.apache.org/} v$2.4.0$, set in cluster mode was used to manage stored data using Apache Hadoop\footnote{https://hadoop.apache.org/}. The various classifier models were implemented and evaluated using Python's Scikit-learn\footnote{https://scikit-learn.org/} library. All scripts were executed within the Jupyter-notebook\footnote{https://jupyter.org/} environment.

\section{Results}\label{sec:result}
Considering the simple classifier \textit{C\_entity} from the first experiment, the Gradient Boosting model yielded a better average score ($61.90\%$ accuracy) and \textit{MCC} ($0.44$) than Random Forest and Adaboost classifiers, as shown in Table \ref{tab:cross-val} (upper section). However, with overall low \textit{MCC} for all classifiers (between $0.22$ and $0.44$), these scores were not sufficient to achieve reliable entities characterization. This led to introducing our cascading machine learning approach, enriching the initial entity dataframe with information gathered from prior classifications in the second experiment.

\begin{table}[!htbp]
\centering
\begin{tabular}{lccccc}
\cline{1-5}
\multicolumn{1}{c}{Model} & Classifier & Score \% & Std \% & MCC  \\ \cline{1-5}
\textit{Adaboost}         & \textit{C\_entity}          & 45.63          & 6.34         &   0.22      \\ 
\textit{Random Forest}     & \textit{C\_entity}          & 59.71          & 1.82         &   0.41     \\ 
\textit{Gradient Boosting}     & \textit{C\_entity}          & 61.90        & 1.36         &   0.44     \\ \cline{1-5}
\textit{Adaboost}         & \textit{C\_final}           & 78.84          & 1.76         &   0.76     \\ 
\textit{Random Forest}         & \textit{C\_final}           & 98.04           & 1.22         &   0.97     \\ 
\textit{Gradient Boosting}     & \textit{C\_final}           & 99.68           & 0.63         &   0.99     \\ \cline{1-5}
\end{tabular}
    \caption{\textit{Average performance of classifiers over five cross-validation repetitions for simple \textit{C\_entity} model (above) and for final model after data enrichment via cascading machine learning \textit{C\_final}}}
    \label{tab:cross-val}
\end{table}

Analyzing the \textit{C\_address}, \textit{C\_motif1} and \textit{C\_motif2} classifiers separately for entity characterization, Table \ref{tab:model_score} shows that outgoing information from the Random Forest classifier resulted to be more accurate than information from Gradient Boosting and Adaboost classifiers (accuracy scores \textgreater$90\%$ for Random Forest). Notably, only using information from the address dataframe, the Random Forest classifier \textit{C\_address} could already achieve an average global accuracy of $\sim96\%$. Due to these results, we only used results obtained from Random Forest classifiers for the subsequent entities dataframe enrichment. Random Forest classifiers not only proved to be the best in terms of accuracy, but also performed with highest speed among the considered classification models.

\begin{table}[!htbp]
\centering
\begin{tabular}{lcccc}
\cline{1-4}
\multicolumn{1}{c}{Model} & \textit{C\_address} \% & \textit{C\_motif1} \% & \textit{C\_motif2} \% \\ \cline{1-4}
\textit{Adaboost}         & 61.54          & 72.69           & 78.27             \\ 
\textit{Random Forest}     & 95.73          & 94.14          & 90.88            \\ 
\textit{Gradient Boosting}     & 83.23          & 83.52         & 83.54            \\ \cline{1-4}
\end{tabular}
    \caption{\textit{Average global C\_address, C\_motif1 and C\_motif2 classifier accuracy calculated via 5-fold cross-validation}}
    \label{tab:model_score}
\end{table}

The final classifiers \textit{C\_final} were fed with the enriched entity dataframe, which comprised the original features from the entity dataframe and included as new features the class predictions obtained from \textit{C\_address}, \textit{C\_motif1} and \textit{C\_motif2} for the respective test data sets. From Table \ref{tab:cross-val} (lower section) it is obvious that the average score result improved significantly by exploiting the information obtained via our cascading approach. Random Forest and Gradient Boosting classifiers again performed better than the Adaboost model, reaching a score of more than $98\%$ (respectively $\sim39\%$ and $\sim38\%$ percentage points higher than the baseline accuracy from \textit{C\_entity}). Furthermore, classification results were more stable during cross-validation, generating low standard deviations between $0.63\%$ and $1.76\%$ and the \textit{MCC} reached values close to $1.0$, relating to close-to-perfect class prediction.

%%%%%%%%%%%%%%%%%%%%%%%%%%%%%%%%%%%%%%%%%%%%%%%%%%%%%%%%%%%%%%%%%%%%%%%%%%%%
%%%%%%%%%%%%%%%%%%%%%%%%%%%%%%%%%%%%%%%%%%%%%%%%%%%%%%%%%%%%%%%%%%%%%%%%%%%%%%
\begin{table*}
  \centering
\begin{tabular}{lcccc|ccc}
\cline{3-8}
                          & \multicolumn{1}{l}{} & \multicolumn{3}{c|}{\textit{C\_entity} model} & \multicolumn{3}{c}{\textit{C\_final} model}         \\ \hline
\multicolumn{1}{c}{Class} & Model                & Precision    & Recall    & F1-score   & Precision & \multicolumn{1}{l}{Recall} & F1-score \\ \hline
\textit{Exchange}         & Adaboost             & 0.51            & 0.68         & 0.57          & 0.77         & 0.78                          & 0.77        \\
\textit{Gambling}         & Adaboost             & 0.22           & 0.14         & 0.17         & 0.75         & 1.00                          & 0.85        \\
\textit{Market}           & Adaboost             & 0.05            & 0.15         & 0.08          & 0.40         & 0.30                          & 0.33       \\
\textit{Mining Pool}             & Adaboost             & 0.20            & 0.16         & 0.17          & 0.11         & 0.2                          & 0.14       \\
\textit{Mixer}            & Adaboost             & 0.69            & 0.78         & 0.71          & 1.00         & 0.98                          & 0.99        \\
\textit{Service}          & Adaboost             & 0.20            & 0.10         & 0.13          & 0.95         & 0.95                          & 0.95        \\ \hline
\textit{Exchange}         & Random Forest             & 0.60            & 0.77         & 0.67          & 0.96         & 1.00                          & 0.98        \\
\textit{Gambling}         & Random Forest             & 0.54            & 0.50         & 0.51          & 1.00           & 1.00                            & 1.00          \\
\textit{Market}           & Random Forest             & 0            & 0         & 0          & 1.00           & 0.85                          & 0.91        \\
\textit{Mining Pool}             & Random Forest             & 0.68            & 0.50         & 0.56          & 1.00         & 0.92                            & 0.96        \\
\textit{Mixer}            & Random Forest             & 0.89            & 0.78         & 0.82          & 1.00           & 1.00                            & 1.00          \\
\textit{Service}          & Random Forest             & 0            & 0         & 0          & 1.00         & 0.93                          & 0.96        \\ \hline
\textit{Exchange}         & Gradient Boosting             & 0.61            & 0.80        & 0.69          & 1.00         & 1.00                          & 1.00       \\
\textit{Gambling}         & Gradient Boosting             & 0.59            & 0.53         & 0.55          & 0.99         & 1.00                          & 0.99        \\
\textit{Market}           & Gradient Boosting             & 0.10            & 0.05         & 0.06          & 1.00         & 1.00                          & 1.00        \\
\textit{Mining Pool}             & Gradient Boosting             & 0.38           & 0.40         & 0.38          & 1.00         & 1.00                          & 1.00        \\
\textit{Mixer}            & Gradient Boosting             & 0.92           & 0.84         & 0.87          & 1.00         & 1.00                          & 1.00        \\
\textit{Service}          & Gradient Boosting             & 0            & 0         & 0          & 1.00         & 0.93                          & 0.96        \\ \hline
\end{tabular}
    \caption{\textit{Average Precision, Recall and F\textsubscript{1}-score calculated in each model implementation for each class}}
    \label{tab:test-metrics}
\end{table*}

In Table \ref{tab:test-metrics}, we present per-class Precision, Recall and F\textsubscript{1}-scores calculated for \textit{C\_entity} (baseline) and \textit{C\_final} (enriched) classifiers for each classification model. Results demonstrate that the simple classifier \textit{C\_entity} - independently of the classification model used - had problems detecting \textit{Service} and \textit{Market} entities (calculated metrics are $0$ or have very low values). However, it is to be noted that these two classes are least represented in terms of distinct entities in the original data set. Random Forest and Gradient Boosting classifiers showed overall good performance in detecting \textit{Mixer} entities for the \textit{C\_entity} approach (F\textsubscript{1}-scores \textgreater$0.8$).

By exploiting the cascading machine learning implementation however, all classifiers improved their classification performance for each class, with most values being close to $1.0$. Only the Adaboost model kept having problems with the classification of \textit{Mining Pool} and \textit{Market} entities. Random Forest and the Gradient Boosting models instead yielded excellent values for Precision, Recall and F\textsubscript{1}-score for each class.

Overall best classification scores were achieved by the \textit{C\_final} implementation with Gradient Boosting models. Data enrichment through prior classification and cascading thus clearly had a highly beneficial impact on classification ability of Gradient Boosting, motivating a further analysis of the importance of individual features from the enriched entity dataframe. We therefore calculated in a next step a feature \textit{importance score} for the enriched entity dataframe.

Generally, the feature importance score provides a score that indicates how useful or valuable each feature was in the construction of the model. The more often an attribute is used to make key decisions, the greater will be its relative importance score. Importance was explicitly calculated through Python's Scikit-learn library for each attribute in the data set, allowing features to be ranked and compared to each other.

\begin{figure}[]
  \centering
    \includegraphics[width=\linewidth]{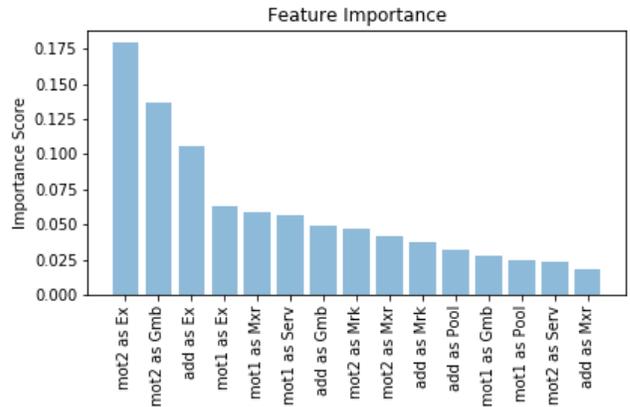}
    \caption{\textit{Top 15 important features from the GB classifier}}
    \label{fig:features}
\end{figure}

Figure \ref{fig:features} shows a list of the top fifteen features for the \textit{C\_final} Gradient Boosting classifier. All fifteen important features were created during the prior classifications taking into account \textit{C\_address}, \textit{C\_motif1} and \textit{C\_motif2}. These features represent how address, 1\_motif, or 2\_motif data, related to a certain entity, were previously classified. This highlights again that the information brought in from prior classifications (first step of the cascade) clearly contributes to much improved entities characterization.

\section{Conclusion and Future work}\label{sec:conclusions}
In this paper, we present a novel approach of how to attack Bitcoin anonymity through entity characterization. Specifically, we demonstrate how a cascading machine learning model combined with an adequate set of input features directly derived from Bitcoin blockchain data (entity and address data) as well as derived via 1\_motif and 2\_motif concepts introduced by Ranshous et al. \cite{ranshous2017exchange} can lead to impressive classification performance for a number of relevant Bitcoin entity classes. In fact, we were able to obtain an average global accuracy score of $99.68\%$ with low standard deviation of $0.63\%$ and a Matthews Correlation Coefficient (\textit{MCC}) of $0.99$ over 5-fold cross validation for a Gradient Boosting model using our cascading approach. 

These final models were indeed able to predict each of the six entity classes used (Exchange, Gambling, Market, Mining Pool, Mixer, Service) with Precision, Recall and F\textsubscript{1}-score values close to $1.0$. Ranshous et al. \cite{ranshous2017exchange} obtained similar results using Random Forest and Adaboost classifiers, however their study was limited to exchange address classification. Jourdan et al. \cite{jourdan2018characterizing} generally obtained lower values for per-class F\textsubscript{1}-score and Precision ranging between $0.67$ and $1.0$ using Gradient Boosting and their approach involved a complex step of model hyper-parameter calibration and required a total number of $315$ input features.

Our approach applies one more classification step in the "classification cascade" generating a set of new entity-related features used for the final classification, but we do not require extensive parameter tuning. Most importantly, we only use $34$ features for the initial classification step (involving address, 1\_motif and 2\_motif) and finally $7$ features from the entities data set plus $3$ x $6 = 18$ new features obtained as outgoing information from the initial classification step. The final classification is thus based on only $25$ features, which equals to less than $10\%$ of features compared to Jourdan et al. Our future work will focus on investigating deeper the matter of feature importance, in order to further reduce the number of relevant features required for obtaining high entity classification performance. This will facilitate the process of attacking Bitcoin anonymity further.

One major drawback of our approach is that we were not able to characterize entities behaving as normal users as this information is not currently available as ground truth data in the WalletExplorer. We had to remove all entities that have not yet been classified in the WalletExplorer from our analysis. Nevertheless, we were able to detect six classes of key Bitcoin services that have previously been associated with illicit financial operations with very high classification scores. We therefore believe that our study can contribute to improving crime investigation and may form a base for developing effective tools assisting law enforcement agencies in uncovering illegal activities within the Bitcoin network.

\section*{ACKNOWLEDGMENT}
This work was partially funded by the European Commission through the Horizon 2020 research and innovation program, as part of the ``TITANIUM" project (grant agreement No 740558).

%%%%%%%%%%%%%%%%%%%%%%%%%%%%%%%%%%%%%%%%%%%%%%%%%%%%%%%%%%%%%%%%%%%%%%%%%%%%%%%%
\addtolength{\textheight}{-12cm} 
\bibliographystyle{splncs04}
\bibliography{main}

\end{document}